\documentclass[a4paper,12pt]{article}
\pdfoutput=1
\usepackage{graphicx, rotating}
\usepackage{hyperref}
\usepackage{slashed}
\usepackage{ifpdf}

\ifx\pdfoutput\undefined
\usepackage[dvips,bookmarks=false]{hyperref}	
\else
\usepackage{hyperref}	
\fi
\hypersetup{colorlinks,bookmarksopen,bookmarksnumbered,citecolor=verdes,
linkcolor=blus,pdfstartview=FitH,urlcolor=rossos}
\def\hhref#1{\href{http://arxiv.org/abs/#1}{#1}} 

\usepackage{amsmath}
\usepackage{slashed}

\newcommand{\beq}{\begin{equation}}
\newcommand{\eeq}{\end{equation}}
\newcommand{\fig}[1]{~\ref{fig:#1}}

\newcount\Mac  \Mac=1  
\newcommand{\ifMac}[2]{\ifnum\Mac=1 #1 \else #2 \fi}
\def\putps(#1,#2)(#3,#4)#5#6{\ifnum\Mac=1 \put(#1,#2){\special{picture #5}}
\else  \put(#3,#4){\includegraphics{#6}} \fi}

\newcommand{\One}{\hbox{1\kern-.24em I}}

\newcommand{\GeV}{\,{\rm GeV}}
\newcommand{\TeV}{\,{\rm TeV}}

\newcommand{\NP}{Nucl. Phys.}

\newcommand{\PL}{Phys. Lett.}
\newcommand{\PR}{Phys. Rev.}

\newcommand{\eq}[1]{~{\rm (\ref{eq:#1})}}

\newcommand{\lascia}[1]{}
\makeatletter
\def\art{\@ifnextchar[{\eart}{\oart}}
\def\eart[#1]#2#3#4#5#6{{\rm #2}, {#3 #4} {\rm (#6) #5} [arXiv:{\hhref{#1}}]}
\def\hepart[#1]#2{{\rm #2, arXiv:\hhref{#1}}}
\newcommand{\oart}[5]{{\rm #1}, {#2 #3} {\rm (#5) #4}}

\newcounter{alphaequation}[equation]
\def\thealphaequation{\theequation\hbox to
0.6em{\hfil\alph{alphaequation}\hfil}}
\def\eqnsystem#1{
\def\@eqnnum{{\rm (\thealphaequation)}}
\def\@@eqncr{\let\@tempa\relax \ifcase\@eqcnt \def\@tempa{& & &} \or
  \def\@tempa{& &}\or \def\@tempa{&}\fi\@tempa
  \if@eqnsw\@eqnnum\refstepcounter{alphaequation}\fi
\global\@eqnswtrue\global\@eqcnt=0\cr}
\refstepcounter{equation} \let\@currentlabel\theequation \def\@tempb{#1}
\ifx\@tempb\empty\else\label{#1}\fi
\refstepcounter{alphaequation}
\let\@currentlabel\thealphaequation
\global\@eqnswtrue\global\@eqcnt=0 \tabskip\@centering\let\\=\@eqncr
$$\halign to \displaywidth\bgroup \@eqnsel\hskip\@centering
$\displaystyle\tabskip\z@{##}$&\global\@eqcnt\@ne
\hskip2\arraycolsep\hfil${##}$\hfil& \global\@eqcnt\tw@\hskip2\arraycolsep
$\displaystyle\tabskip\z@{##}$\hfil
\tabskip\@centering&\llap{##}\tabskip\z@\cr}

\def\endeqnsystem{\@@eqncr\egroup$$\global\@ignoretrue} \makeatother

\def\circa#1{\,\raise.3ex\hbox{$#1$\kern-.75em\lower1ex\hbox{$\sim$}}\,}

\usepackage{multicol}
\usepackage{color}
\definecolor{rosso}{cmyk}{0,1,1,0.4}
\definecolor{rossos}{cmyk}{0,1,1,0.55}
\definecolor{rossoc}{cmyk}{0,1,1,0.2}
\definecolor{blu}{cmyk}{1,1,0,0.3}
\definecolor{blus}{cmyk}{1,1,0,0.6}
\definecolor{bluc}{cmyk}{1,1,0,0.1}
\definecolor{verde}{cmyk}{0.92,0,0.59,0.25}
\definecolor{verdec}{cmyk}{0.92,0,0.59,0.15}
\definecolor{verdes}{cmyk}{0.92,0,0.59,0.4}
\definecolor{grigio}{cmyk}{0,0,0,0.07}
\definecolor{rosa}{cmyk}{0,0.1,0.1,0.02}
\definecolor{rosino}{cmyk}{0,0.05,0.05,0.02}
\definecolor{rosas}{cmyk}{0,0.3,0.25,0.05}
\definecolor{celeste}{cmyk}{0.1,0,0,0.02}
\definecolor{giallino}{cmyk}{0,0,0.4,0.02}
\definecolor{rosso}{cmyk}{0,1,1,0.4}
\definecolor{rossos}{cmyk}{0,1,1,0.55}
\definecolor{rossoc}{cmyk}{0,1,1,0.2}
\definecolor{blu}{cmyk}{1,1,0,0.3}
\definecolor{bluc}{cmyk}{1,1,0,0.1}
\definecolor{blucc}{cmyk}{0.7,0.5,0,0}
\definecolor{viola}{cmyk}{0,1,0,0.6}
\definecolor{viola2}{cmyk}{0,1,0.2,0.6}
\definecolor{verde}{cmyk}{0.92,0,0.59,0.25}
\definecolor{verdec}{cmyk}{0.92,0,0.59,0.15}
\definecolor{verdes}{cmyk}{0.92,0,0.59,0.4}
\definecolor{verdino}{cmyk}{0.12,0,0.09,0.05}
\definecolor{giallo}{cmyk}{0,0,1,0}
\definecolor{gialloverde}{cmyk}{0.44,0,0.74,0}

\font\tenrsfs=rsfs10 at 12pt

\font\sevenrsfs=rsfs7
\font\fiversfs=rsfs5
\newfam\rsfsfam
\textfont\rsfsfam=\tenrsfs
\scriptfont\rsfsfam=\sevenrsfs
\scriptscriptfont\rsfsfam=\fiversfs
\def\mathscr#1{{\fam\rsfsfam\relax#1}}

\begin{document}

\color{black}
\vspace{0.5cm}
\begin{center}
{\Huge\bf\color{rossos}The fine-tuning price\\ of the early LHC}\\
\bigskip\color{black}\vspace{0.6cm}
{{\large\bf Alessandro Strumia}
} \\[7mm]
{\it Dipartimento di Fisica dell'Universit{\`a} di Pisa and INFN, Italia}\\[3mm]
{\it  National Institute of Chemical Physics and Biophysics, Ravala 10, Tallin, Estonia}\\[3mm]
\end{center}
\bigskip
\centerline{\large\bf\color{blus} Abstract}
\begin{quote}\large
LHC already probed and excluded most of the parameter space of the
Constrained Minimal Supersymmetric Standard Model
allowed by previous experiments. 
Only about $0.3\%$ of the CMSSM parameter space survives.
This fraction rises to about $0.9\%$ if the bound on the Higgs mass can be circumvented.

\color{black}
\end{quote}


\section{Introduction}
When LEP started 20 years ago, the main topic of high-energy physics was 
finding the right supersymmetric unified model, its embedding in string theory and
understanding how it predicts a zero cosmological constant.

But supersymmetry was not found and LEP opened a ``little hierarchy problem''~\cite{GRS,paradox}:
experimental bounds on sparticle masses made difficult to fully solve the higgs mass hierarchy problem.
Furthermore, cosmological observations strongly suggested a small but non-zero cosmological constant~\cite{cosmo},
opening a new hierarchy problem with no known solution.

\smallskip

Now LHC is starting and its most fundamental goal is telling why the 
the weak scale is much below the Planck scale: is it small due to some natural reason
(as many theorists expect) or for some other ``unnatural'' reason
(such as anthropic selection in a multiverse)?

Supersymmetry remains the main candidate natural solution to the weak scale hierarchy problem.
The CMS and ATLAS collaborations published the first results of searches for
supersymmetric particles
looking at events with jets and missing energy, without and with a lepton,
first in data at $\sqrt{s}=7\TeV$ with $35$/pb of integrated luminosity~\cite{LHC35/pb}, later increased to about $1.1$/fb~\cite{LHC1/fb}.

Supersymmetry was not found in such  LHC data, and the LHC collaborations produced bounds
in the CMSSM model:  degenerate squarks and gluinos must be typically heavier than up to $1.1\TeV$ (see fig.\fig{scan} below).
This is significantly stronger than previous bounds.


\smallskip

Again, the main implications of such negative experimental searches concerns naturalness,
which is the heart of the main question: is the weak scale naturally small?

\begin{figure}
$$\includegraphics{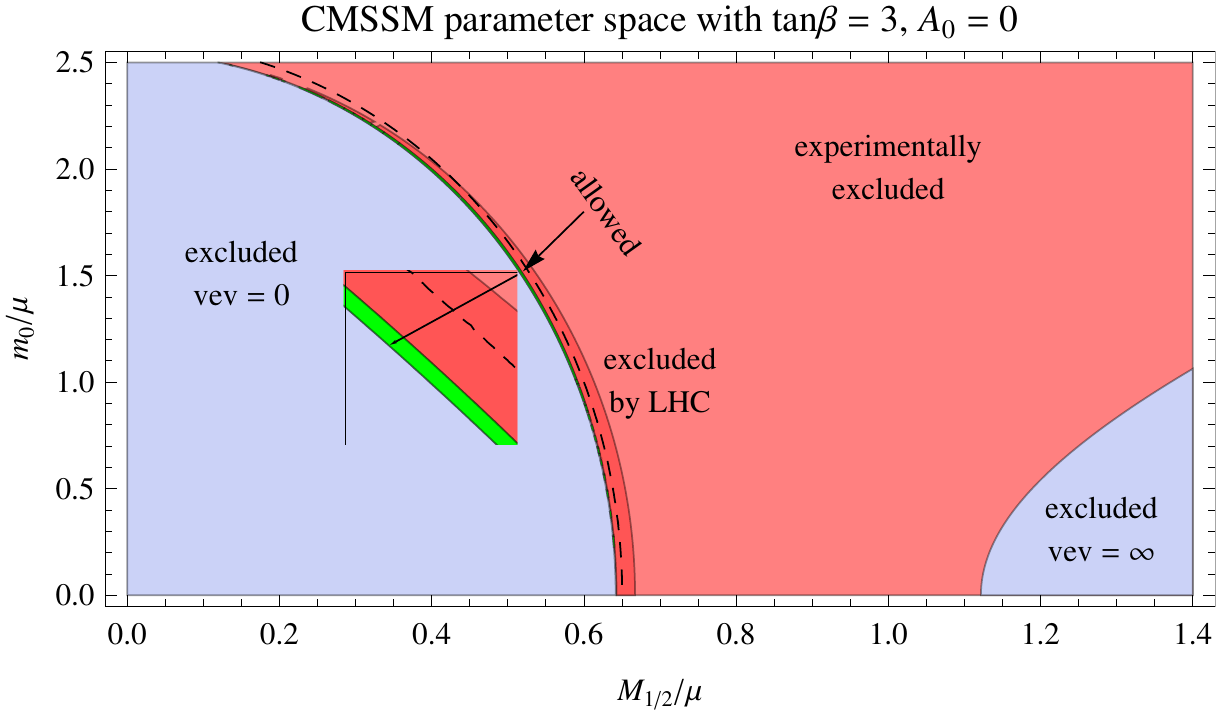}$$
\caption{\em A typical example of the parameter space of the CMSSM model. The green region is allowed (see it in the enlarged box).
The dashed line around the boundary of the allowed region is the prediction of the model considered in~\cite{BS}.
\label{fig:RR}}
\end{figure}

\begin{figure}[t]
$$\includegraphics[width=0.45\textwidth]{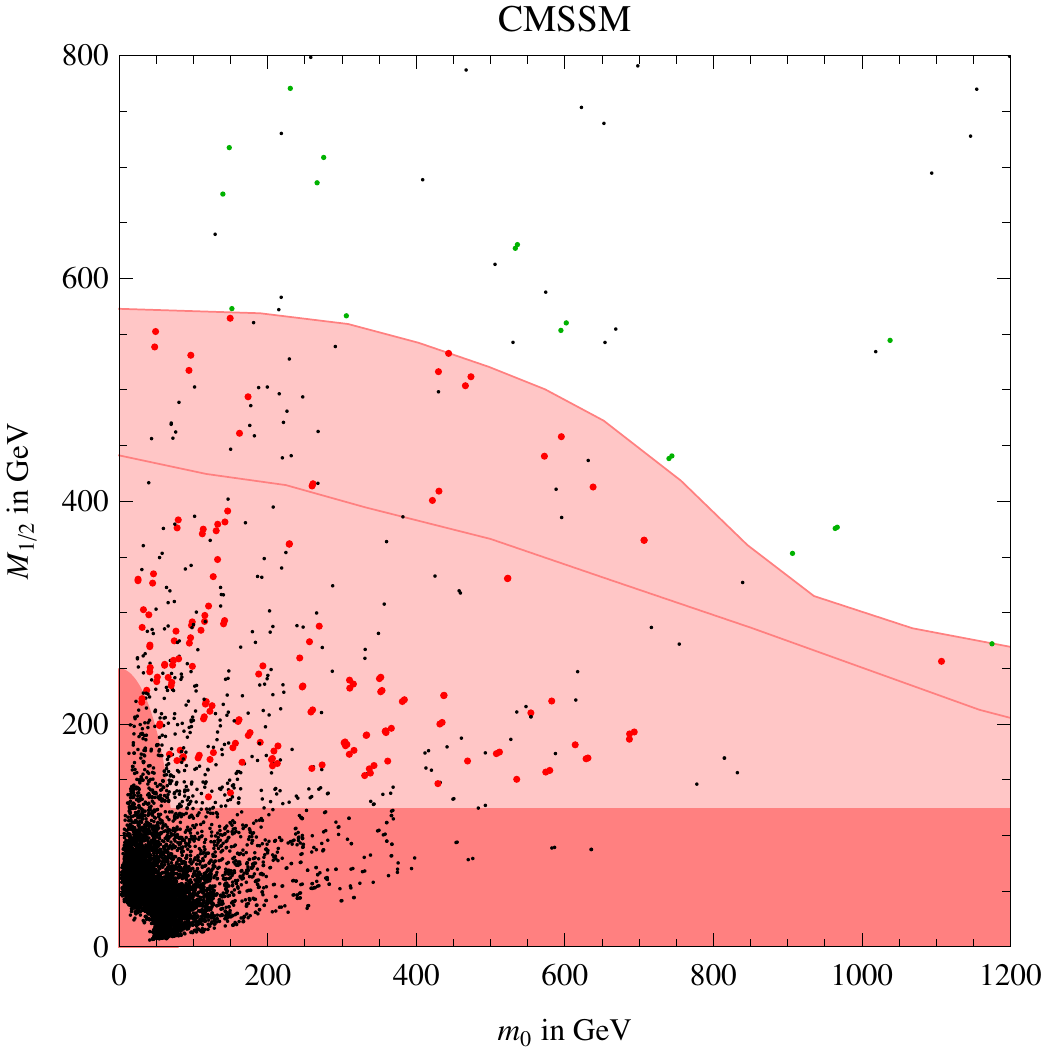}\qquad
\includegraphics[width=0.45\textwidth]{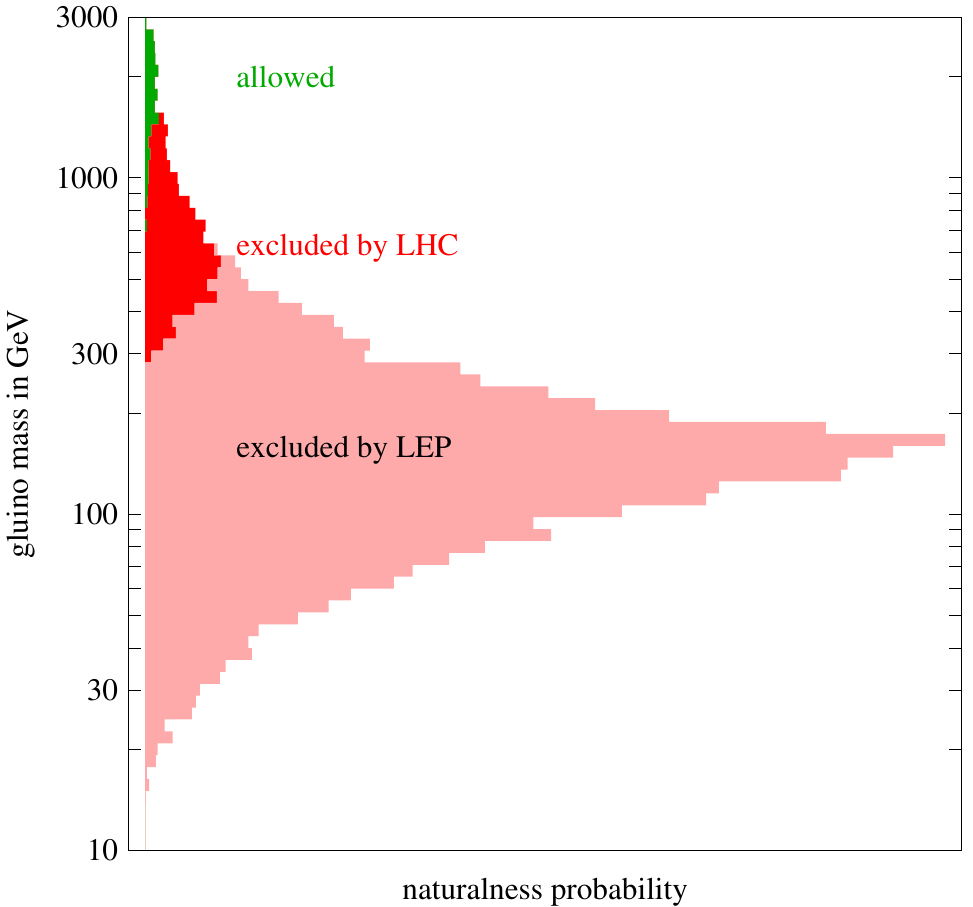}$$
\caption{\em Left: naturalness scan of the CMSSM.
Red points are excluded by LHC, black points have been excluded earlier,
green points are allowed. The darker pink region was excluded by LEP and the pink region  by early LHC
(the red lines show the various bounds from ATLAS and CMS).
Right: ``naturalness probability distribution'' for the gluino mass in the CMSSM.  Only its tail was allowed after LEP, and the tail of the tail remains allowed after first LHC data.
\label{fig:scan}}
\end{figure}

\section{Naturalness}
To illustrate the naturalness problem of the CMSSM model we recall that it predicts the $Z$ mass to be
\begin{equation}
M^2_Z \approx  0.7 M_{3}^2+0.2 m^2_0 -2 \mu^2= (91\GeV)^2\times  100(\frac{M_3}{1.1\TeV})^2+\cdots
\label{eq:MZ}
\end{equation}
where  $M_3\approx 2.6 M_{1/2}$ is the gluino mass, 
$M_{1/2}$ and $m_0$ are the unified gaugino and scalar masses at the unification scale;
the $\mu$ term is renormalized at the weak scale, and $\cdots$ denotes the $m_0^2$ and $\mu^2$ terms.
We here assumed $\tan\beta=3$ and $A_0=0$,
such that the top Yukawa coupling renormalized at the unification scale is $\lambda_t(M_{\rm GUT})\approx 0.5$.
Eq.\eq{MZ} means that the natural sparticle scale is $M_{1/2}\sim m_0\sim \mu\sim M_Z$
and that an accidental cancellation by a part in $\approx 100$
is needed if $M_3>1.1\TeV$.

\smallskip

Eq.\eq{MZ} can be used to fix the overall SUSY mass scale, such that the CMSSM model  has
two free adimensional  parameters: the ratios $M_{1/2}/\mu$ and $m_0/\mu$
($\tan\beta=3$ and $A_0=0$ are for the moment kept fixed).
Such parameter space is plotted in fig.\fig{RR}:
\begin{itemize}
\item The light-gray regions are theoretically excluded because the minimum of the potential is not the physical one:
in the left region one would have $M_Z^2<0$ which means that the true minimum is at $v=0$;
in the bottom-right region the potential is unstable when the two higgses have equal vev.
\item The red region in the middle is theoretically allowed, but has now been experimentally excluded.
The darker red shows the new region probed and excluded by LHC with respect to the previous
LEP bounds, approximated to be $M_2>100\GeV$.
\item
The green region is allowed. Indeed it is
close to the boundary where $M_Z=0$ and thereby has $M_Z\ll m_0,M_{1/2},\mu$.
\end{itemize}
The smallness of the allowed region is a manifestation of the ``little hierarchy problem''.

\medskip

We now relax the restriction on $A_0$ and $\tan\beta$ (or equivalently $B_0$) and study naturalness
proceeding along the lines of~\cite{GRS}, as briefly summarized below.

We randomly scan the full theoretically allowed adimensional parameters of the model
(the adimensional ratios between $m_0$, $M_{1/2}$, $\mu$, $A_0$, $B_0$ as well as the top Yukawa coupling
$\lambda_t$,
all renormalized at the unification scale)
determining the overall SUSY mass scale and $\tan\beta$ from the potential minimization condition.
Thanks to the last step, we sample the full CMSSM parameter space according to its natural density
(rare accidental cancellations that make sparticles heavy
happen rarely). We compute how rare
are the still allowed sparticle spectra, as in~\cite{GRS}
that claimed that only $5\%$ of the CMSSM parameter space survived to LEP.

More precisely we perform the following scan
\beq\label{eq:s1}
m_0=(\frac{1}{3^2}\div 3)_{\rm log} m_{\rm SUSY},\qquad
|\mu_0|,M_{1/2}=(\frac{1}{3}\div 3)_{\rm log} m_{\rm SUSY},\qquad
A_0,B_0=(-3\div 3)_{\rm lin} M_{1/2}
\eeq
and verify that it gives results similar to other possibilities such as
\beq\label{eq:s2}
m_0,|\mu_0|,M_{1/2}=(\frac{1}{3}\div 3)_{\rm log} m_{\rm SUSY},\qquad
A_0,B_0=(-3\div 3)_{\rm lin} m_0
\eeq
or as
\beq\label{eq:s3} m_0,|\mu_0|,M_{1/2},|B_0|,|A_0|=(0\div 1)_{\rm lin}m_{\rm SUSY}.\eeq 
where the pedices `lin' and `log' respectively denote a flat probability distribution in linear or logarithmic scale
within the given range.

More formally, this is a Monte Carlo Bayesian technique that starts with an arbitrary
non-informative prior probability density function (implicitly defined by the `random scans' in eq.s\eq{s1} to\eq{s3})
and gives a set of points in parameter space
with probability density roughly equal to the inverse
of the various fine-tuning measures proposed to approximate the naturalness issue~\cite{FT}.
The above procedure makes no use of any fine-tuning parameter, and automatically
takes into account all fine-tunings: not only the one needed to have $M_Z \ll m_0, M_{1/2},\mu$, but also 
the one needed to have $\tan\beta\gg1$,
or the fine-tuning on $\lambda_t$ that can give a small or even negative $m_0^2$ coefficient in eq.\eq{MZ},
such that the $M_3^2$ term can be cancelled by $m_0^2$ rather than by $\mu^2$.
The scanning is restricted to top quark masses within $3$ standard deviations of the present measured value,
$m_t = (173.1\pm1.1)\GeV$~\cite{mt}.

\begin{table}
$$\begin{array}{c|ccc}
\hbox{experimental} & \multicolumn{3}{c}{\hbox{fraction of surviving CMSSM parameter space}}\\ 
\hbox{bound} &\hbox{any $m_h$} & m_h >100\GeV & m_h >110\GeV\\  \hline
\hbox{LEP} & 10\% & 4\% & 1\%\\
\hbox{LHC} & 0.9\% &0.5\% & 0.3\% \\
\end{array}$$
\caption{\label{tab}\em Fraction of the CMSSM parameter space that survives to the various bounds.}
\end{table}

\smallskip

A technical detail.
The MSSM minimization equations generalize eq.\eq{MZ} taking into account one loop corrections to the potential.
To understand their relevance, we recall that
at tree level the higgs mass is predicted to be $m_h^{\rm tree}\le M_Z\cos2\beta$, while at loop level it can be above
the experimental limit $m_h>114\GeV$.  
The effect of minimizing the one loop potential (rather than the tree level potential)
is essentially equivalent to rescaling the overall SUSY mass scale by a factor $m_h/m_h^{\rm tree}$, which helps naturalness.

\medskip

We consider the three main bounds on sparticles, that can be roughly summarized as follows:
\begin{itemize}
\item[$1)$] The LHC bound on (mainly) the gluino and squark
masses is plotted in fig.\fig{scan}a.\footnote{The CMS and ATLAS collaborations computed bounds for $\tan\beta=3$ or 10 and $A_0=0$~\cite{LHC35/pb,LHC1/fb}.
The dominant bound from events with jets and missing energy is essentially a bound on the gluino and squark masses,
so that the dependence on $A_0$ and $\tan\beta$ can be neglected.
The subleading  bound from events with one lepton has only a moderate dependence on $\tan\beta$, that we also ignore.}
We find that this bound alone excludes about $99\%$ of the CMSSM parameter space.

\item[$2)$] LEP tells that all charged sparticles (charginos, sleptons, stops...)  are heavier than about 100 GeV,
unless they are quasi-degenerate with the lightest supersymmetric particle.
Such bounds alone excluded about $90\%$ of the CMSSM parameter space~\cite{GRS}.

\item[$3)$] The LEP bound on the Higgs mass ($m_h>114\GeV$ in the SM) is potentially even stronger but it is not robust and
deserves a dedicated discussion.
\end{itemize}
As well known, the bound on the Higgs mass can exclude the whole MSSM, 
because the MSSM  predicts at tree level a higgs lighter than $M_Z$ and at loop
level a higgs lighter than about $125\GeV$.
The precise value logarithmically depends on the sparticle mass scale;
we compute it using our own code and using the more precise {\sc SoftSusy} code~\cite{SoftSUSY}.
However, one can modify the MSSM to increase the predicted higgs mass (e.g.\ adding a singlet as in the NMSSM),
avoiding the fine-tuning price of the higgs mass bound (and alleviating the whole fine tuning~\cite{Barb}).
Furthermore, in regions of the MSSM parameter space with $\mu\sim M_Z$ and large $\tan\beta$
the higgs coupling to the $Z$ is reduced and
a weaker bound $m_h\circa{>}100\GeV$ applies.
More generically, the bound on the higgs mass can be weakened (down to about 100 GeV)
modifying the theory such that
the higgs has new dominant decay modes which are more difficult to see experimentally, or which have not been experimentally studied~\cite{ALEPH}.  
Summarizing, many models can  alleviate the fine-tuning problem related to the higgs mass.

\medskip

The motivation for such models gets now washed out by the LHC bound, which has nothing to do with the higgs mass.
Table~\ref{tab} shows that naturalness remains a problem, even if one can circumvent the bound on the higgs mass.


In its last column we stick to the CMSSM and approximate the LEP bound $m_h>114\GeV$ as $m_h^{\rm th}>110\GeV$, where
$m_h^{\rm th}$ is the higgs mass as computed by state-of-the-art codes~\cite{SoftSUSY},
that have a theoretical uncertainty estimated to be about $\pm3\GeV$.
With such a bound the allowed fraction of the CMSSM parameter space decreases down to about $0.3\%$. 
The other scans (eq.\eq{s2} or eq.\eq{s3}) would give lower comparable fractions of allowed parameter space.


\begin{figure}[t]
$$\includegraphics[width=0.45\textwidth]{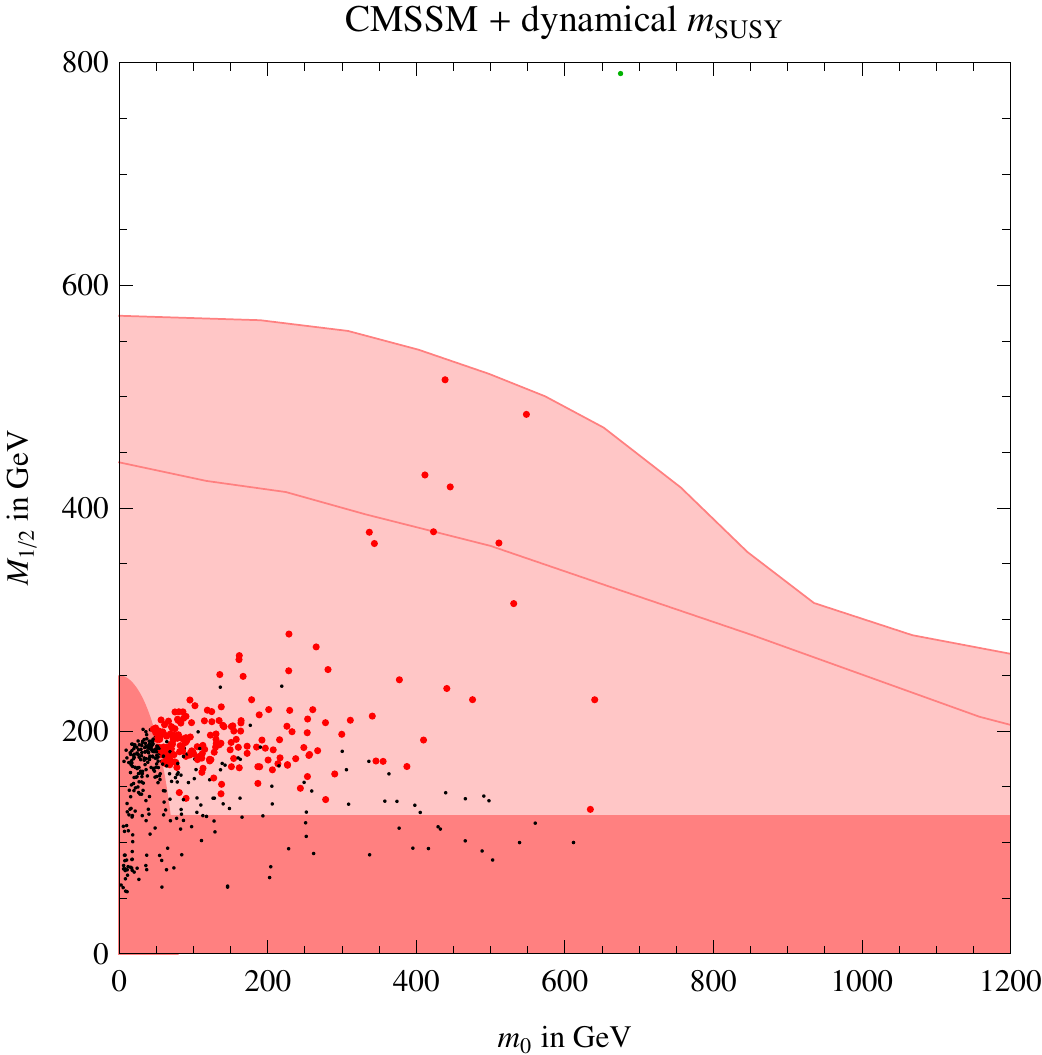}$$
\caption{\em As in the previous figure\fig{scan}a, but assuming
that the sparticle mass scale is dynamically determined by minimizing the MSSM potential~\cite{BS},
such that it is naturally heavier than $M_Z$.
\label{fig:scanslide}}
\end{figure}

\section{Conclusions}
LHC data sharpen the ``little hierarchy problem'' of supersymmetry.
Various tentative solutions were proposed.
The simplest solution, a gluino lighter than what predicted by unification~\cite{Kane,GRS}, 
seems now excluded by the new LHC bound.

It is maybe useful to see the problem in this way:
in SUSY models renormalization effects from the Planck scale down to the weak scale
can radiatively break the weak symmetry, making the determinant of the squared higgs mass term negative
below some scale $Q_0$ which in principle is anyway between the weak and Planck scales:
the little hierarchy problem means that breaking of the weak gauge group
happens at the last moment, $Q_0\sim v$~\cite{paradox}.

Therefore, one possible interpretation is that the Higgs is the pseudo-Goldstone boson of some symmetry broken
at some scale $Q_0$
around the weak scale; however concrete models often look less plausible than the fine-tuning they avoid~\cite{pseudo}.

Another possibility considered in~\cite{BS} is assuming that the overall scale of soft supersymmetry breaking is dynamically fixed
by minimizing the {\em weak part only} of the potential (this assumption looks 
implausible, as recognized in~\cite{BS}),
such that the SUSY scale must be just below $Q_0$;
the precise computation gives a neat prediction of the form
$m_{\rm SUSY}\approx 4\pi M_Z/12$,
where $4\pi$ is a loop factor and the factor 12 counts the spins and colors in the most relevant diagram~\cite{BS}.
Applied to the CMSSM model such prediction  is
plotted as dashed line in fig.\fig{RR}: it lies around the present allowed/excluded border.
Fig.\fig{scanslide} shows the naturalness scan in the full parameter space: we see that
even in this case LHC probed most of the parameter space.
LEP tested SUSY masses around the $Z$ mass, and now LHC reached the next milestone, testing SUSY masses a loop
factor above the $Z$ mass.

\smallskip

Maybe the weak scale is small due to anthropic selection, and attempts of 
keeping it technically small are like attempts of dragging the \ae{}ther.
The scenario of~\cite{BS} was reconsidered in~\cite{NEW} with a different motivation: the authors 
imagine a supersymmetric
multiverse where for some unknown reason
the weak scale is ``almost never''  broken ($Q_0\ll m_{\rm SUSY}$), such that $Q_0\sim m_{\rm SUSY}$
is ``more likely''.  The qualitative expectation is similar to the prediction of~\cite{BS}, but 
$m_{\rm SUSY}$ can be made heavier by order one factors making the ``almost never'' stronger,
at the price of making MSSM vacua more rare than the SM with an unnaturally light higgs.


 
 \bigskip
 
All these beautiful ideas and
the history of the Michelson-Morley experiment teach us that a negative experimental search can have deep theoretical implications.
\begin{flushright}
``{\em History repeats itself, first as tragedy, second as farce}''.
Karl Marx
\end{flushright}

\small

\paragraph{Acknowledgements}  The author thanks Ben Allanach,
Riccardo Rattazzi and Andrea Romanino for useful
communications.
This work was supported by the ESF grant MTT8 and by SF0690030s09 project.
\small



\begin{thebibliography}{nn}


\bibitem{GRS} \art[hep-ph/9811386]{L. Giusti, A. Romanino, A. Strumia}{\NP}{B550}{3}{1999}.
See also  \hepart[hep-ph/9904247]{A. Strumia}.


\bibitem{paradox}
\hepart[hep-ph/0007265]{R. Barbieri, A. Strumia}.

\bibitem{cosmo}
WMAP collaboration,  arXiv:1001.4744 and references therein.

\bibitem{LHC35/pb}
\hepart[1101.1628]{CMS collaboration}.
\hepart[1102.5290]{ATLAS collaboration}.
\hepart[1102.2357]{ATLAS collaboration}.

\bibitem{LHC1/fb}  LHC bounds on supersymmetry with more than 1/fb of integrated luminosity
have been presented at the EPS-HEP conference on 25/7/2011
(web site \url{http://eps-hep2011.eu}) by G. Tonelli (for the CMS collaboration) and by D. Charlton (for the ATLAS collaboration),
as well as in various parallel talks.






%


\bibitem{FT}
\art{R. Barbieri and G.F. Giudice}{\NP}{B306}{63}{1988}.
\art{B. de Carlos and J.A. Casas}{\PL}{B309}{320}{1993}.
\art{G.W. Anderson and D.J. Casta\~no}{\PL}{B347}{300}{1995}\art{}{\PR}{D52}{1995}{1693}.
\art[hep-ph/9611204]{P. Ciafaloni and A. Strumia}{\NP}{B494}{41}{1997}.
\art[hep-ph/9712234]{P.H. Chankowski, J. Ellis and S. Pokorski}{\PL}{B423}{327}{1998}.
\hepart[hep-ph/9808275]{P.H. Chankowski, J. Ellis, M. Olechowski and S. Pokorski}.


\bibitem{mt}
Tevatron Electroweak Working Group and CDF and D0 Collaboration,
  arXiv:0903.2503.
  
  \bibitem{SoftSUSY} B.C. Allanach, Comput. Phys. Commun. 143 (2002) 305 [arXiv:hep-ph/0104145].
  
\bibitem{Barb}
R.~Barbieri, L.~J.~Hall, Y.~Nomura {\it et al.},
  Phys.\ Rev.\  {D75}, 035007 (2007)
  [hep-ph/0607332].
  
  

\bibitem{ALEPH}
For recent experimental effort, see
 ALEPH collaboration,
  JHEP {1005} (2010)  049
  [arXiv:1003.0705].


%
\bibitem{Kane}
  \hepart[hep-ph/9801449]{D. Wright}.
G.L.~Kane, S.F.~King,
  Phys.\ Lett.\  {B451}, 113 (1999)
  [hep-ph/9810374].
  






\bibitem{pseudo}
  Z.~Berezhiani, P.~H.~Chankowski, A.~Falkowski {\it et al.},
  Phys.\ Rev.\ Lett.\  {96}, 031801 (2006).
  [hep-ph/0509311].
\art[hep-ph/0510294]{C.~Csaki, G.~Marandella, Y.~Shirman, A. Strumia}{\PR}{D73}{035006}{2006}.
S.~Chang, L.~J.~Hall, N.~Weiner,
  Phys.\ Rev.\  {D75}, 035009 (2007).
  [hep-ph/0604076].
B.~Bellazzini, S.~Pokorski, V.~S.~Rychkov {\it et al.},
  JHEP {0811}, 027 (2008).
  [arXiv:0805.2107].
B.~Bellazzini, C.~Csaki, A.~Delgado {\it et al.},
  Phys.\ Rev.\  {D79}, 095003 (2009).
  [arXiv:0902.0015].




\bibitem{BS}
\art[hep-ph/0005203]{R. Barbieri, A. Strumia}{\PL}{B490}{247}{2000}.





%



\bibitem{NEW}
G. Giudice and R. Rattazzi,   Nucl.\ Phys.\  {B757} 19 (2006)   [hep-ph/0606105].





%











\end{thebibliography}
\end{document}